\documentstyle[12pt]{article}
\begin{document}
\title{\huge\bf Extended axial model}

\author{R. Amorim\thanks{\noindent Electronic mail: amorim @ if.ufrj}
and J. Barcelos-Neto\thanks{\noindent Electronic mails: ift03001 @
ufrj and barcelos @ vms1.nce.ufrj.br}\\ Instituto de F\'{\i}sica\\
Universidade Federal do Rio de Janeiro\\ RJ 21945-970 - Caixa Postal
68528 - Brasil}

\maketitle
\abstract
We study an extension of the axial model where local gauge symmetries
are taken into account. The anomaly of the axial current is
calculated by the Fujikawa formalism and the model is also solved.
Besides the well known features of the particular models (axial and
Schwinger) it was obtained an effective interaction of scalar and
gauge fields via a topological current.

\vfill
\vspace{1cm}
\newpage

\section{Introduction}

\bigskip
The axial model corresponds to a theory where a real scalar field
interacts with a fermionic axial current via a derivative coupling.
It was introduced by Rothe and Stamatescu almost
twenty years ago \cite{RS} and has been studied since then on its
more diverse aspects \cite{B,DH,FV,BF}. Its Lagrangian density reads
\footnote{We adopt throughout this paper the following convention and
notation: $\eta_{\mu\nu}={\rm diag.} (+1,-1)$, $\epsilon^{01}=+1$,
$\{\gamma^\mu,\gamma^\nu\}=2\eta^{\mu\nu}$,
$\gamma^\mu=\gamma^0\gamma^{\mu\dagger}\gamma^0$
($\gamma^{0\dagger}=\gamma^0$, $\gamma^{1\dagger}=-\gamma^1$),
$\gamma_5=\gamma^0\gamma^1$ ($\gamma_5^\dagger=\gamma_5$),
$\gamma^\mu\gamma^\nu=\eta^{\mu\nu}+\epsilon^{\mu\nu}\gamma_5$,
$\gamma_5\gamma^\mu=\epsilon^{\mu\nu}\gamma_\nu$ (these last two are
only true in two dimensions).}

\begin{equation}
\label{1}
{\cal L}=i\,\bar\psi\gamma^\mu\,
\bigl(\partial_\mu+ig_0\gamma_5\partial_\mu\phi\bigr)\,
\psi+{\cal L}_\phi\,,
\end{equation}

\bigskip\noindent
where

\begin{equation}
\label{2}
{\cal L}_\phi=\frac{1}{2}\,
\Bigl(\partial_\mu\phi\partial^\mu\phi
-m_0^2\,\phi^2\Bigr)\,.
\end{equation}

\bigskip\noindent
We easily observe that it exhibits global gauge and chiral-gauge
symmetries. In consequence, from the Noether's theorem, both vector
and axial currents are conserved. The two main features of this model
are

\bigskip
({\it i}) The divergence of the axial current is anomalous:

\begin{equation}
\label{3}
\partial_\mu j_5^\mu=\frac{i\,g_0}{\pi}\,
{\sqcap\!\!\!\!\sqcup}\,\phi\,,
\end{equation}

\bigskip\noindent
where $j^\mu_5$ is defined as

\begin{equation}
\label{4}
j^\mu_5=\bar\psi\gamma^\mu\gamma_5\psi\,.
\end{equation}

\bigskip
({\it ii}) The mass $m_0$ is renormalized to

\begin{equation}
\label{5}
m^2=\frac{m_0^2}{1-\strut\frac{\displaystyle g_0^2}
{\displaystyle\pi}}\,.
\end{equation}

\bigskip
This model does not have the corresponding local gauge symmetries.
Apparently, one possibility of having them would be to consider the
coupling of the axial current with a gauge field $A_\mu$, namely,
$g_0\bar\psi\gamma^\mu\psi\,A_\mu$, and taking
$A_\mu=\partial_\mu\phi$ as a particular gauge choice. However, this
cannot be true. A simple argument against this procedure is that the
coupling term above does not have the correct dimension.

\medskip
The consistent way of implementing local gauge transformations in the
axial model is actually by means of a gauge field, but taken
independently of $\partial_\mu\phi$. The general Lagrangian density
then reads

\begin{equation}
\label{6}
{\cal L}=i\,\bar\psi\not\!\! D\psi
-g_0\bar\psi\gamma^\mu\gamma_5\psi\,\partial_\mu\phi
+{\cal L}_\phi+{\cal L}_A\,,
\end{equation}

\bigskip\noindent
where ${\cal L}_A$ is the well known gauge field term

\begin{equation}
\label{7}
{\cal L}_A=-\frac{1}{4}\,F_{\mu\nu}\,F^{\mu\nu}\,,
\end{equation}

\bigskip\noindent
and the covariant derivative reads

\begin{equation}
\label{8}
D_\mu=\partial_\mu-ieA_\mu\,.
\end{equation}

\bigskip\noindent
We list below the mass dimension of the quantities that appear in the
Lagrangian~(\ref{6}):

\begin{equation}
\label{9}
[\psi]={1\over2}\,,\hskip.5cm
[\phi]=0\,,\hskip.5cm
[A_\mu]=0,\hskip.5cm
[e]=1\,,\hskip.5cm
[g_0]=0\,,\hskip.5cm
[m_0]=1\,.\hskip.5cm
\end{equation}

\bigskip
The model described by (\ref{6}) is in fact a mixing of the axial
model, described by expressions (\ref{1}) and (\ref{2}), and the
Schwinger model \cite{Sch}. The purpose of this work is to study the
features of such extended model, which will be called ``extended
axial model" (EAM). It might be opportune to first make some comments
about the EAM:

\bigskip
({\it i}) The derivative which is acting on the scalar field cannot be
replaced by a covariant one. This is so because the scalar field is
real and consequently does not couple to the electromagnetic one (the
gauge transformation of $\phi$ is zero).

\bigskip
({\it ii)} Inadvertently, we could think to make $g_0=e$ to
eliminate the second term of (\ref {6}) by a gauge transformation of
$A_\mu$. As it was previously said, this cannot be done since $g_0$
and $e$ do not have the same mass dimensions.

\bigskip
({\it iii)} Another wrong reasoning would be the trial to obtain the
scalar term $-{1\over2}\,\partial_\mu\phi\partial^\mu\phi$ from the
gauge field one $-{1\over4}\,F_{\mu\nu}F^{\mu\nu}$ and vice-versa by
considering that in two-dimensions the gauge field can always be
written as

\begin{equation}
\label{10}
A_\mu=\partial_\mu\sigma
+\epsilon_{\mu\nu}\partial^\nu\rho\,.
\end{equation}

\bigskip\noindent
Also here, this would not be succeeded.  $\sigma$ and $\rho$ have
mass-dimensions $-1$ and, consequently, cannot be related to $\phi$.
In fact, in case of replacing $A_\mu$ given by (\ref{10}) into
(\ref{7}), we would obtain terms in
${\sqcap\!\!\!\!\sqcup}\,\sigma\,{\sqcap\!\!\!\!\sqcup}\,\sigma$,
${\sqcap\!\!\!\!\sqcup}\,\rho\,{\sqcap\!\!\!\!\sqcup}\,\rho$ etc.

\bigskip
So we conclude that Lagrangian (\ref{6}) cannot be simplified. It has
to be used in the way it appears. We shall see that the anomaly of
the axial current of the EAM is more general and contains crossed
terms of both sectors. However, the renormalized mass of the scalar
field and the mass acquired by the photon field are precisely the
same of the usual models when taken separately. A point to be
emphasized is that after solving the EAM, the scalar field
effectively interacts with the gauge one via a topological current.

\medskip
Our paper is organized as follows. In Sec. 2 we calculate the anomaly
of the axial current of the EAM by means of the Fujikawa path
integral technique~\cite{Fu}. The model is solved by integrating on
the fermionic fields in Sec. 3. Some final remarks are left for Sec.
4.

\section{Axial current anomaly}
\bigskip
To calculate the anomaly of the axial current we make use of the
Fujikawa technique~\cite{Fu}. We begin by writing down the general
expression of the vacuum functional

\begin{equation}
\label{11}
Z=N\int[d\bar\psi][d\psi][d\phi][dA]\,\exp\,\,\{i\,S\}\,,
\end{equation}

\bigskip\noindent
where $S$ is the action corresponding to the Lagrangian density
(\ref{6}) plus some gauge-fixing term
\footnote{It is understood that we have functionally integrated on
ghost fields that come from the Faddeev-Popov method.}.
The chiral anomaly arises in the path integral formalism from the
fact that the measure $[d\bar\psi][d\psi]$ is not invariant under
chiral gauge transformations. In the infinitesimal case, these
transformations read

\begin{eqnarray}
\psi^\prime(x)&=&
\bigl[1+i\,\epsilon(x)\,\gamma_5\bigr]\,\psi(x)\,,\nonumber\\
\bar\psi^\prime(x)&=&
\bar\psi(x)\bigl[1+i\,\epsilon(x)\,\gamma_5\bigr]\,,
\label{12}
\end{eqnarray}

\bigskip\noindent
and we obtain \cite{Fu}

\begin{equation}
\label{13}
[d\bar\psi][d\psi]=[d\bar\psi^\prime][d\psi^\prime]\,
\exp\,\,\Bigl\{2i\int d^2x\,\,\epsilon(x)\,I(x)\Bigr\}\,,
\end{equation}

\bigskip\noindent
where

\begin{equation}
\label{14}
I(x)=\sum_n\phi^\dagger_n(x)\,\gamma_5\,\phi_n(x)\,.
\end{equation}

\bigskip\noindent
The quantities $\phi_n(x)$ form a complete and orthonormal set of
eigenfunctions of some Hermitian operator $O_H$. To determine the
anomaly, it is then necessary to perform the calculation of the sum
above. However, it is well known that sums like
$\sum_n\phi_n^\dagger(x)\Gamma_{\mu\nu\dots}\phi(y)$, where
$\Gamma_{\mu\nu\dots}$ are product of gamma matrices, are divergent
when $x=y$. The way found by Fujikawa to regularize these sums is to
introduce an exponential factor in order to avoid contributions of
big eigenvalues. Concentrating on our particular case, this is done
in the following way

\begin{eqnarray}
I(x)&=&\lim_{M\rightarrow\infty}
\sum_n\phi^\dagger_n(x)\,\gamma_5\,\phi_n(x)\,
{\rm e}^{-\lambda^2_n/M^2}\,,\nonumber\\
&=&\lim_{M\rightarrow\infty}
\sum_n\phi^\dagger_n(x)\,\gamma_5\,
{\rm e}^{-O^2_H/M^2}\,\phi_n(x)\,,\nonumber\\
&=&\lim_{M\rightarrow\infty\atop x\rightarrow y}\,
\Bigl(\gamma_5\,{\rm e}^{-O^2_H/M^2}\Bigr)_{ij}\,
\sum_n\phi_{nj}(x)\,\phi_{ni}^\dagger(y)\,,\nonumber\\
&=&\lim_{M\rightarrow\infty\atop x\rightarrow y}\,
\Bigl(\gamma_5\,{\rm e}^{-O^2_H/M^2}\Bigr)_{ij}\,
\delta_{ij}\,\delta^{(2)}(x-y)\,,\nonumber\\
&=&\lim_{M\rightarrow\infty\atop x\rightarrow y}\,
{\rm tr}\,\,\gamma_5\,{\rm e}^{-O^2_H/M^2}\,\delta^{(2)}(x-y)\,.
\label{15}
\end{eqnarray}

\bigskip\noindent
According to the Fujikawa method, the operator that has to be used to
regularize the sum is the one that appears in the theory. In our
case, we have

\begin{equation}
\label{16}
\tilde{\not\!\!D}=\gamma^\mu\,\Bigl(\partial_\mu
-ieA_\mu-ig_0\gamma_5\partial_\mu\phi\Bigr)\,.
\end{equation}

\bigskip
The point is that this operator is not Hermitian. An usual procedure
used in literature is to go to the Euclidean space. We do it by
letting $x^0\rightarrow-ix_4$, $\partial_0\rightarrow i\partial_4$,
$\gamma^0\rightarrow i\gamma_4$,
$\gamma^\mu\partial_\mu\rightarrow-\gamma_\mu\partial_\mu$. In this
space, all gamma matrices are anti-Hermitians
\footnote{In the Euclidean space, our initial convention and notation
have to be changed to $\{\gamma_\mu,\gamma_\nu\}=-2\delta_{\mu\nu}$,
$\gamma_\mu\gamma_\nu=-\delta_{\mu\nu}-i\epsilon_{\mu\nu}\gamma_5$,
$\gamma_5\gamma_\mu=i\epsilon_{\mu\nu}\gamma_\nu$,
$\gamma_5=i\gamma_1\gamma_4$, $\epsilon_{14}=+1$ etc.}.
So, the terms
$\gamma_\mu\partial_\mu$ and $ie\gamma_\mu A_\mu$ are Hermitians, but
$ig_0\gamma_\mu\gamma_5\partial_\mu\phi$ is not. Consequently, the
operator

\begin{equation}
\label{17}
\not\!\!{\tilde D}_E=\gamma_\mu\,\Bigl(\partial_\mu
-ieA_\mu-ig_0\gamma_5\partial_\mu\phi\Bigr)\,,
\end{equation}

\bigskip\noindent
obtained from

\begin{equation}
\label{18}
\not\!\!{\tilde D}\longrightarrow-\not\!{\tilde D}_E
\end{equation}

\bigskip\noindent
is still not Hermitian.

\bigskip
There is a simple way of circumventing this problem. It consists in
taking a kind of analytical extension of the field $\phi$ when we go
to the Euclidean space \cite{FV,AB}. So, instead of the operator
(\ref{17}) we use the Hermitian one

\begin{equation}
\label{19}
\not\!\!{\tilde D}_H=\gamma_\mu\,\Bigl(\partial_\mu
-ie\,A_\mu+g_0\gamma_5\partial\tilde\phi\Bigr)\,,
\end{equation}

\bigskip\noindent
where it was taken

\begin{equation}
\label{20}
\phi\longrightarrow i\tilde\phi\,.
\end{equation}

\bigskip
We shall use $\not\!\!{\tilde D}_H$ as the regulating operator of the
Fujikawa technique. It is then just a matter of algebraic calculation
to obtain

\begin{equation}
\label{21}
I(x)=\frac{1}{2\pi}\,\Bigl(g_0\,{\sqcap\!\!\!\!\sqcup}\,\tilde\phi
-2ieg_0\,A_\mu\partial_\mu\tilde\phi
+e\,\epsilon_{\mu\nu}\,\partial_\mu A_\nu\Bigr)\,.
\end{equation}

\bigskip\noindent
Rotating back to the Minkowski space and not forgetting to make
$i\tilde\phi\rightarrow\phi$, we have

\begin{equation}
\label{22}
I(x)=\frac{1}{2\pi}\,\Bigl(ig_0\,{\sqcap\!\!\!\!\sqcup}\phi
+2eg_0\,A^\mu\partial_\mu\phi
+e\,\epsilon^{\mu\nu}\partial_\mu A_\nu\Bigr)\,.
\end{equation}

\bigskip\noindent
With this result the measure changes to

\begin{equation}
\label{23}
[d\bar\psi][d\psi]=[d\bar\psi^\prime][d\psi^\prime]\,
\exp\,\,\Bigl\{\frac{i}{\pi}\int d^2x\,\,
\epsilon(x)\,\Bigl(ig_0\,{\sqcap\!\!\!\!\sqcup}\,\phi
+2eg_0\,A^\mu\partial_\mu\phi
+e\,\epsilon^{\mu\nu}\partial_\mu A_\nu\Bigr)\Bigr\}\,,
\end{equation}

\bigskip\noindent
which implies through

\begin{equation}
\label{24}
\lim_{\epsilon\rightarrow0}\,
\frac{\delta Z}{\delta\epsilon(x)}=0
\end{equation}

\bigskip\noindent
that
\begin{equation}
\label{25}
\partial_\mu\,j^\mu_5=-\frac{1}{2\pi}\,
\Bigl(e\,\epsilon_{\mu\nu}\,F^{\mu\nu}
+2ig_0\,{\sqcap\!\!\!\!\sqcup}\,\phi
+4eg_0\,A^\mu\partial_\mu\phi\Bigr)\,.
\end{equation}

\bigskip\noindent
We notice that when we take $e=0$, we obtain the anomaly of the axial
current of the Rothe and Stamatescu's model. The corresponding
anomaly of the Schwinger model is obtained by taking $g_0=0$. We
emphasize the presence of a mixing term with $eg_0$ coupling. This
leads to an effective interaction between $A_\mu$ and $\phi$. We are
going to discuss the consequences of this fact with more details in
the next section.

\section {Path integral solution of the model}
\bigskip
Both in Schwinger and axial models there are effective theories that
can be obtained by integrating on the fermionic fields. Let us see
what kind of effective theory can be obtained here. Considering the
fermionic part of the Lagrangian (\ref{6}) and taking the $A_\mu$
field just as

\begin{equation}
\label{26}
A_\mu=\epsilon_{\mu\nu}\,\partial^\nu\rho\,,
\end{equation}

\bigskip\noindent
that corresponds to take the gauge condition where $\sigma=0$ (see
expression \ref{10}), we have

\begin{eqnarray}
{\cal L}_F&=&i\,\bar\psi\,\gamma^\mu\,\bigl(\partial_\mu
-ie\,\epsilon_{\mu\nu}\,\partial^\nu\rho\bigr)\,\psi
+g_0\,\bar\psi\,\gamma_5\,\gamma^\mu\psi\,\partial_\mu\phi\,,\nonumber\\
&=&i\,\bar\psi\,\gamma^\mu\,
\bigl(\partial_\mu-i\,\gamma_5\,\partial_\mu\xi\bigr)\,\psi\,,
\label{27}
\end{eqnarray}

\bigskip\noindent
where

\begin{equation}
\label{28}
\xi=e\,\rho-g_0\,\phi\,.
\end{equation}

\bigskip\noindent
We notice that if we choose the gauge transformation

\begin{equation}
\label{29}
\psi={\rm e}^{i\gamma_5\xi}\,\chi\,,
\end{equation}

\bigskip\noindent
we will obtain that the coupling of the fermionic current with
$\partial_\mu\rho$ disappears. Doing this iteratively by means of
infinitesimal gauge transformations we get for the measure
\cite{DH,FV,Fu}

\begin{equation}
\label{30}
[d\bar\psi][d\psi]=[d\bar\chi][d\chi]\,
\exp\,\,\Bigl(-\,\frac{i}{2\pi}\,\int d^2x\,
\partial_\mu\xi\,\partial^\mu\xi\Bigr)\,.
\end{equation}

\bigskip\noindent
Replacing the expression of $\xi$ given by (\ref{28}), we get

\begin{eqnarray}
[d\bar\psi][d\psi]&=&[d\bar\chi][d\chi]\,
\exp\,\Bigl[-\frac{i}{2\pi}\,\int d^2x\,
\Bigl(e^2\,\partial_\mu\rho\partial^\mu\rho
-2eg_0\partial_\mu\rho\partial^\mu\phi\nonumber\\
&&\phantom{[d\bar\chi][d\chi]\,
\exp\,\Bigl[-\frac{i}{2\pi}\,\int d^2x\,
\Bigl(e^2\,\partial_\mu\rho\partial^\mu\rho}
+g_0^2\,\partial_\mu\phi\partial^\mu\phi\Bigr)\Bigr]\,,\nonumber\\
&=&[d\bar\chi][d\chi]\,
\exp\,\Bigl[-\frac{i}{2\pi}\,\int d^2x\,
\Bigl(e^2\,A_\mu A^\mu
-2eg_0\epsilon^{\mu\nu}\partial_\nu\phi\,A_\mu\nonumber\\
&&\phantom{[d\bar\chi][d\chi]\,
\exp\,\Bigl[-\frac{i}{2\pi}\,\int d^2x\,
\Bigl(e^2\,\partial_\mu\rho\partial^\mu\rho}
-g_0^2\,\partial_\mu\phi\partial^\mu\phi\Bigr)\Bigr]\,.
\label{31}
\end{eqnarray}

\bigskip\noindent
The vacuum functional then turns to be

\begin{eqnarray}
Z=N\int[d\bar\chi][d\chi][dA][d\phi]\,
\exp\,\Bigl\{i\int d^2x\,
\Bigl[i\,\chi\not\!\partial\chi
\!\!&-&\!\!\frac{1}{4}\,F_{\mu\nu}\,F^{\mu\nu}
+\frac{e^2}{2\pi}\,A_\mu A^\mu\nonumber\\
\!\!&+&\!\!\frac{1}{2}\,\Bigl(1-\frac{g^2_0}{\pi}\Bigr)\,
\partial_\mu\phi\,\partial^\mu\phi
-m_0^2\phi^2\nonumber\\
\!\!&-&\!\!\frac{eg_0}{\pi}\,\epsilon^{\mu\nu}\,
\partial_\nu\phi\,A_\mu\Bigr]\Bigr\}\,.
\label{32}
\end{eqnarray}

\bigskip
As one observes, the resulting fermionic field does not interact with
$A_\mu$ and $\phi$ anymore. It may disappear from the theory by
integrating over it and absorbing the result into the renormalization
factor $N$. The so obtained effective Lagrangian contains the well
known results of the Schwinger and the axial model, that is to say,
the photon field acquires a mass given by $e/\sqrt\pi$ and the mass
$m_0$ of the scalar field is renormalized to
$m_0\,(1-g_0^2/\pi)^{-1/2}$. The new point here is that the scalar
field, that initially did not interact with the gauge field, now does
via the current

\begin{equation}
\label{33}
J^\mu=\frac{eg_0}{\pi}\,\epsilon^{\mu\nu}\,
\partial_\nu\phi\,.
\end{equation}

\bigskip
As can be trivially verified, $J^\mu$ is conserved independently of
the equation of motion and it is not associated to any symmetry of
the action. These facts give a topological character to $J^\mu$. Due
to the boundary conditions we are using in the evaluation of most
the quantities throughout the work (fields vanishing at the spatial
infinity), there is only a null topological charge associated to
$J^\mu$.

\section{Conclusion}
In this brief report we have considered the quantization of the
extended axial model, which contains the Schwinger and the axial
models as convenient limits. By using the Fujikawa prescription, it
was possible to show that the axial current is actually anomalous.
This anomaly has the usual contributions from the axial model sector
as well as from the Schwinger model, but it contains also a crossed
term which is a new feature of the considered model, expressing the
fact that at the quantum level there is an interference of both
sectors.

\medskip
This fact also comes true when the model is solved by functionally
integrating on the fermionic sector. The effective Lagrangian which
survives presents a coupling between the EM field and the scalar one
by means of a topological current.

\medskip
We note that as the coupling between the gauge field and the scalar
one is done by a bilinear term, further integrations on $A_\mu$ or
$\phi$ could be done, leading to effectively pure scalar or pure
vectorial theories with non local kinetical terms.

\vskip 1cm
\noindent {\bf Acknowledgment:} This work is supported in part by
Conselho Nacional de Desenvolvimento Cient\'{\i}fico e Tecnol\'ogico
- CNPq (Brazilian Research Agency).

\end{document}